\documentclass[conference]{IEEEtran}
\usepackage{booktabs}
\usepackage{color}
\usepackage{graphicx}
\ifCLASSINFOpdf
\else
\fi
\begin{document}

\title{Timbre Analysis of Music Audio Signals with Convolutional Neural Networks}

\author{\IEEEauthorblockN{Jordi Pons, Olga Slizovskaia, Rong Gong, Emilia G\'{o}mez and Xavier Serra}
\IEEEauthorblockA{Music Technology Group, Universitat Pompeu Fabra, Barcelona.\\
name.surname@upf.edu
}}

\maketitle

\begin{abstract}
The focus of this work is to study how to efficiently tailor Convolutional Neural Networks (CNNs) towards learning timbre representations from log-mel magnitude spectrograms. We first review the trends when designing CNN architectures. Through this literature overview we discuss which are the crucial points to consider for efficiently learning timbre representations using CNNs. From this discussion we propose a design strategy meant to capture the relevant time-frequency contexts for learning timbre, which permits using domain knowledge for designing architectures. In addition, one of our main goals is to design efficient CNN architectures -- what reduces the risk of these models to over-fit, since CNNs' number of parameters is minimized.
Several architectures based on the design principles we propose are successfully assessed for different research tasks related to timbre: singing voice phoneme classification, musical instrument recognition and music auto-tagging.
\end{abstract}

\IEEEpeerreviewmaketitle

\section{Introduction}

Our goal is to discover novel deep learning architectures that can efficiently model music signals,  what  is  a  very  challenging undertaking.
After showing that is possible to design efficient CNNs \cite{ponsdesigning1} for modeling temporal features --tempo \& rhythm-- we now focus on studying how to efficiently learn timbre\footnote{See first paragraph of Section II for a formal definition of timbre.} representations, one of the most salient musical features. 

Music audio processing techniques for timbre description can be divided in two groups: \textit{(i)} \textit{bag-of-frames} methods and \textit{(ii)} methods based on \textit{temporal modeling}. On the one hand, \textit{bag-of-frames} methods have shown to be limited as they just model the statistics of the frequency content along several frames \cite{porter2015acousticbrainz}. On the other hand, methods based on \textit{temporal modeling}  consider the temporal evolution of frame-based descriptors \cite{rabiner1989tutorial}\cite{roebel2015automatic} -- some of these methods are capable of representing spectro-temporal patterns, that can model the temporal evolution of timbre \cite{roebel2015automatic}. Then, for example, attack-sustain-release patterns can be jointly represented.

Most previous methodologies --either based on \textit{(i)} or \textit{(ii)}-- require a dual pipeline: first, descriptors need to be extracted using a pre-defined algorithm and parameters; and second, (temporal) models require an additional framework tied on top of the proposed descriptors -- therefore, descriptors and (temporal) models are typically not jointly designed. 
Throughout this study, we explore modeling timbre by means of deep learning with the input set to be magnitude spectrograms. This \textit{quasi} end-to-end learning approach allows minimizing the effect of the fixed pre-processing steps described above. 
Note that no strong assumptions over the descriptors are required since a generic perceptually-based pre-processing is used: log-mel magnitude spectrograms. 
Besides, deep learning can be interpreted as a temporal model (if more than one frame is input to the network) that allows learning spectro-temporal descriptors from spectrograms (\textit{i.e.} with CNNs in first layers).
In this case, learnt descriptors and temporal model are jointly optimized, what might imply an advantage when compared to previous methods.

From the different deep learning approaches, we focus on CNNs due to several reasons: \textit{(i)} by taking spectrograms as input, one can interpret filter dimensions in time-frequency domain; and \textit{(ii)} CNNs can efficiently exploit invariances --such as time and frequency invariances present in spectrograms-- by sharing a reduced amount of parameters.
We identified two general trends for modeling timbre using spectrogram-based CNNs: using \textit{small-rectangular filters} ($m \ll M$ and $n \ll N$)\footnote{CNNs input is set to be log-mel spectrograms of dimensions $M{\times}N$ and the CNN filter dimensions to be $m{\times}n$. \textit{M} and \textit{m} standing for the number of frequency bins and \textit{N} and \textit{n} for the number of time frames.} \cite{choi2016automatic}\cite{han2017deep} or using \textit{high filters} ($m \leq M$ and $n \ll N$)$^2$ \cite{dieleman2014end}\cite{lee2009unsupervised}.

$\cdot$ \textit{\underline{Small-rectangular filters}} inquire the risk of limiting the representational power of the first layer since these filters are typically \textit{too small} for modeling spread spectro-temporal patterns \cite{ponsdesigning1}. Since these filters can only represent sub-band characteristics (with a small frequency context: $m \ll M$) for a short period of time (with a small time context: $n \ll N$) these can only learn, for example: onsets or bass notes \cite{pons2016experimenting}\cite{choi2015auralisation}. But these filters might have severe difficulties on learning cymbals' or snare-drums' time-frequency patterns in the first layer since such a spread context can not fit inside a small-rectangular filter.\footnote{Section II further expands this discussion with more details.}

$\cdot$ Although \textit{\underline{high filters}} can fit most spectral envelopes, these might end up with a lot of weights to be learnt from (typically small) data -- risking to \textit{over-fit} and/or to \textit{fit noise}. See Fig. 1 (right) for two examples of filters fitting noise as a result of having available more context than the required for modeling onsets and harmonic partials, respectively.$^3$

Additionally, most CNN architectures use unique filter shapes in every layer \cite{choi2016automatic}\cite{dieleman2014end}\cite{han2017deep}. However, recent works point out that using different filter shapes in each layer is an efficient way to exploit CNN's capacity \cite{ponsdesigning1}\cite{phan2016robust}. For example, Pons \textit{et al.} \cite{ponsdesigning1} proposed using different musically motivated filter shapes in the first layer to efficiently model several musically relevant time-scales for learning temporal features. In Section II we propose a novel approach to this design strategy which facilitates learning musically relevant time-frequency contexts while minimizing the risk of noise-fitting and over-fitting for timbre analysis. Out of this design strategy, several CNN models are proposed. Section III assesses them for three research tasks related to timbre: singing voice phoneme classification, musical instrument recognition and music auto-tagging.

\section{CNNs design strategy for timbre analysis}

Timbre is considered as the \textit{``color"} or the \textit{``quality"} of a sound \cite{wessel1979timbre}. It has been found to be related to the spectral envelope shape and to the time variation of spectral content \cite{peeters2011timbre}.
Therefore, it is reasonable to assume timbre to be a time-frequency expression and then, \textbf{\textit{magnitude spectrograms}} are an adequate \textbf{\textit{input}}. Although phases could be used, these are not considered -- this is a common practice in the literature \cite{choi2016automatic}\cite{dieleman2014end}\cite{han2017deep}, and this investigation focuses on how to exploit the capacity of spectrograms to represent timbre.
Moreover, timbre is often defined by what it is not: \textit{``a set of auditory attributes of sound events in addition to pitch, loudness, duration, and spatial position"} \cite{mcadams2013musical}. Then, we propose ways to design CNN architectures invariant to these attributes:

$\mathbf{\cdot}$ \textbf{\textit{Pitch invariance}}. By enabling filters to convolve through the frequency domain of a mel spectrogram (\textit{a.k.a.} $f_0$ shifting), the resulting filter and feature map can represent timbre and pitch information separately. However, if filters do not capture the whole spectral envelope encoding timbre --because these model a small frequency context--, previous discussion does not necessarily hold.
Additionally, depending on the used spectrogram representation (\textit{i.e.} STFT or mel) CNN filters might be capable of learning more robust pitch invariant features. Note that STFT timbre patterns are $f_0$ dependent. However, mel timbre patterns are more pitch invariant than STFT ones because these are based in a different (perceptual) frequency scale.
Besides, a deeper representation can be pitch invariant if a max-pool layer spanning all over the vertical axis\footnote{\textit{N'} and \textit{M'} denote, in general, the dimensions of any feature map. Therefore, although the filter map dimensions will be different depending on the filter size, we refer to these dimensions by the same name: \textit{N'} and \textit{M'}.} of the feature map (\textit{M'}) is applied to it: \textit{MP(M',}$\cdot$).

$\mathbf{\cdot}$ \textbf{\textit{Loudness invariance}} for CNN filters can be approached by using weight decay -- L2-norm regularization of filter weights. By doing so, filters are normalized to have low energy and energy is then expressed into feature maps. Loudness is a perceptual term that we assume to be correlated with energy.

$\mathbf{\cdot}$  \textbf{\textit{Duration invariance}}. Firstly, $m{\times}1$ filters are time invariant by definition since these do not capture duration. Temporal evolution is then represented in the feature maps. Secondly, sounds with determined length and temporal structure (\textit{i.e.} kick drums or cymbals) can be well captured with $m{\times}n$ filters. These are also duration invariant because such sounds last a fixed amount of time. Note the resemblance between first layer $m{\times}1$ filters with frame-based descriptors; and between first layer $m{\times}n$ filters with spectro-temporal descriptors.

$\mathbf{\cdot}$ \textbf{\textit{Spatial position invariance}} is achieved by down-mixing (\textit{i.e.} averaging all channels) whenever the dataset is not mono.

\begin{figure}[h]
\centering
\includegraphics[width=0.40\paperwidth]{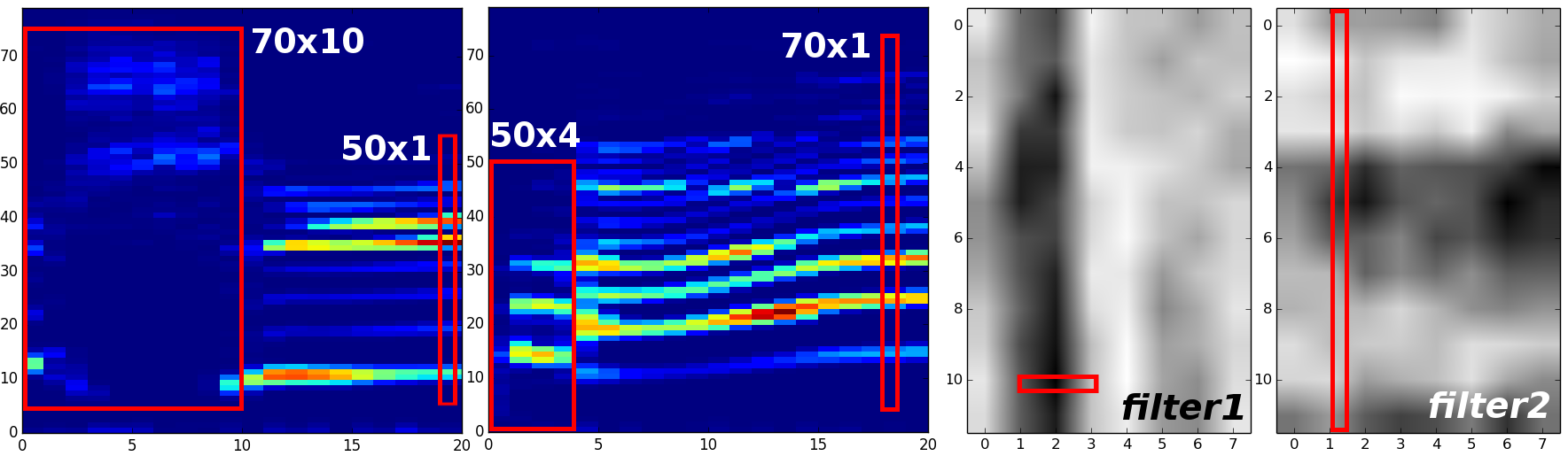}
\caption{\textbf{Left}: two spectrograms of different sounds used for the singing voice phoneme classification experiment. \textbf{Right}: two trained small-rectangular filters of size $12{\times}8$. Relevant time-frequency contexts are highlighted in red.}
\label{specs_filters}
\end{figure}

From previous discussion, we identify the filter shapes of the first layer to be an important design decision -- they play a crucial role for defining pitch invariant and duration invariant CNNs. For that reason, we propose to use \textbf{\textit{domain knowledge for designing filter shapes}}. For example, by visually inspecting Fig. \ref{specs_filters} (left) one can easily detect the relevant time-frequency contexts in a spectrogram: frequency $\in [50,70]$ and time $\in [1,10]$ -- which can not be efficiently captured with several small-rectangular filters. These measurements provide an intuitive guidance towards designing efficient filter shapes for the first CNN layer.

Finally, we discuss how to efficiently learn timbre features with CNNs.
Timbre is typically expressed at different scales in spectrograms -- \textit{i.e.} cymbals are more spread in frequency than bass notes, or vowels typically last longer than consonants in singing voice. If a unique filter shape is used within a layer, one can inquire the risk of: \textit{(a)} fitting noise because too much context is modeled 
and/or \textit{(b)} not modeling enough context. 

$\mathbf{\cdot}$ \underline{\textit{Risk \textit{(a)}}}. Fig. \ref{specs_filters} (right) depicts two filters that have fit noise. Observe that \textit{filter1} is repeating a noisy copy of an onset throughout the frequency axis, and \textit{filter2} is repeating a noisy copy of three harmonic partials throughout the temporal axis. Note that much more efficient representations of these musical concepts can be achieved by using different filter shapes: $1{\times}3$ and $12{\times}1$, respectively (in red). Using the adequate filter shape allows minimizing the risk to fit noise and the risk to over-fit the training set (because the CNN size is also reduced).

$\mathbf{\cdot}$ \underline{\textit{Risk \textit{(b)}}}. The frequency context of \textit{filter2} is too small to model the whole harmonic spectral envelope, and it can only learn three harmonic partials -- what is limiting the representational power of this (first) layer. A straightforward solution for this problem is to increase the frequency context of the filter. However note that if we increase it too much, such filter is more prone to fit noise. Using different filter shapes allows reaching a compromise between risk \textit{(a)} and \textit{(b)}.

\textbf{\textit{Using different filter shapes within the first layer}} seems crucial for an efficient learning with spectrogram-based CNNs. This design strategy allows to efficiently model different musically relevant time-frequency contexts. Moreover, this design strategy ties very well with the idea of using the available domain knowledge for designing filter shapes -- that can intuitively guide the different filter shapes design so that spectro-temporal envelopes can be efficiently represented within a single filter.
Note that another possible solution might be to combine several filters (either in the same layer or going deep) until the desired context is represented. However, several reasons exist for supporting the here proposed approach: \textit{(i)} the Hebbian principle from neuroscience  \cite{hebb1949organization}: \textit{``cells that fire together, wire together"}, and \textit{(ii)} learning complete spectro-temporal patterns within a single filter allows to inspect and interpret the learnt filters in a compact way.

Above discussion introduces the \textbf{\textit{fundamentals}} (in bold italics) of the proposed design strategy for timbre analysis.

\section{Experiments}
Audio is fed to the network using fixed-length log-mel spectrograms. Phases are discarded. Spectrograms are normalized: zero mean and variance one. Activation functions are ELUs \cite{clevert2015fast}. Architectures are designed according to the proposed strategy and previous discussion -- by employing: weight decay regularization, monaural signals, and different filter shapes in the first layer. Each network is trained optimizing the cross-entropy with SGD from random initialization \cite{HeZR015}. The best model in the validation set is kept for testing.

In the following, we assess the validity of the proposed design strategy with 3 general tasks based on timbre modeling:

\subsection{Singing voice phoneme classification}

The jingu\footnote{\textit{``Jingu"} is also known as \textit{``Beijing opera"} or \textit{``Peking opera"}.} a cappella singing audio dataset used for this study \cite{black2014automatic} is annotated with 32 phoneme classes\footnote{Annotation and more details can be found in:\\\phantom~~~~~~~~~~~~~https://github.com/MTG/jingjuPhonemeAnnotation} and consists of two different role-types of singing: \textit{dan} (young woman) and \textit{laosheng} (old man). The \textit{dan} part has 42 recordings (89 minutes) and comes from 7 singers; the \textit{laosheng} part has 23 arias (39 minutes) and comes from other 7 \textit{laosheng} singers. Since the timbral characteristics of \textit{dan} and \textit{laosheng} are very different, the dataset is divided in two. Each part is then randomly split --train (60\%), validation (20\%) and test (20\%)-- for assessing the presented models for the phoneme classification task. Audio was sampled at 44.1 kHz. STFT was performed using a 25ms window (2048 samples with zero-padding) with a hop size of 10ms.
This experiment assesses the feasibility of taking architectural decisions based on domain knowledge for an efficient use of the network's capacity in small data scenarios. The goal is to do efficient deep learning by taking advantage of the design strategy we propose.
This experiment is specially relevant because, in general, no large annotated music datasets are available -- this dataset is an example of this fact. The proposed architecture has a single wide convolutional layer with filters of various sizes. Input is of size $80{\times}21$ -- the network takes a decision for a frame given its context: $\pm$10ms, 21 frames in total. We use $128$ filters of sizes $50{\times}1$ and $70{\times}1$, $64$ filters of sizes $50{\times}5$ and $70{\times}5$, and $32$ filters of sizes $50{\times}10$ and $70{\times}10$ -- considering the discussion in section II. A max-pool layer of $2{\times}N'$ follows before the 32-way softmax output layer with 30\% dropout. \textit{MP(2,N')} was chosen to achieve time-invariant representations while keeping the frequency resolution. 

We use overall classification accuracy as evaluation metric and results are presented in Table \ref{table:performance_am}. As a baseline, we also train a 40-component Gaussian Mixture Models (GMMs), a fully-connected MLP with 2 hidden layers (MLP) and Choi \textit{et al.}'s architecture \cite{choi2016automatic}, that is a 5-layer CNN with small-rectangular filters of size $3{\times}3$ (Small-rectangular). All architectures are adapted to have a similar amount of parameters so that results are comparable. GMMs features are: 13 coefficients MFCCs, their deltas and delta-deltas. $80{\times}21$ log-mel spectrograms are used as input for the other baseline models. Implementations are available online\footnote{https://github.com/ronggong/EUSIPCO2017}.

\begin{table}[!ht]
\centering
\caption{Models performance for \textit{dan} and \textit{laosheng} datasets.}
\label{table:performance_am}
\begin{tabular}{lcccc}
\toprule
 & \textit{dan} / \#param & \textit{laosheng} / \#param \\
 \midrule
Proposed & \textbf{0.484} / 222k & \textbf{0.432} / 222k\\
Small-rectangular & 0.374 / 222k & 0.359 / 222k \\
GMMs & 0.290 / - & 0.322 / - \\
MLP & 0.284 / 481k & 0.282 / 430k \\
 \bottomrule
\end{tabular}
\end{table}

Proposed architecture outperforms other models by a significant margin (although being a single-layer model), what denotes the potential of the proposed design strategy. Deep models based on small-rectangular filters --which are state-of-the-art in other datasets \cite{choi2016automatic}\cite{han2017deep}-- do not perform as well as the proposed model in these small datasets. As future work, we plan to investigate deep models that can take advantage of the richer representations learnt by the proposed model.

\subsection{Musical instrument recognition}

IRMAS \cite{bosch2012} training split contains 6705 audio excerpts of 3 seconds length labeled with a single predominant instrument. Testing split contains 2874 audio excerpts of length 5$\sim$20 seconds labeled with more than one predominant instrument. 11 pitched class instruments are annotated. Audios are sampled at 44.1kHz. The state-of-the-art for this dataset corresponds to a deep CNN based on small-rectangular filters (of size $3{\times}3$) by Han \textit{et al.} \cite{han2017deep}. Moreover, another baseline is provided based on a standard bag-of-frames approach + SVM classifier proposed by Bosch \textit{et al.} \cite{bosch2012}. We experiment with two architectures based on the proposed design strategy:

$\cdot$ \textit{Single-layer} has a single but wide convolutional layer with filters of various sizes. The input is set to be of size $96{\times}128$. We use 128 filters of sizes $5{\times}1$ and $80{\times}1$, 64 filters of sizes $5{\times}3$ and $80{\times}3$, and 32 filters of sizes $5{\times}5$ and $80{\times}5$. We also max-pool the \textit{M'} dimension to learn pitch invariant representations: MP(\textit{M',16}). 50\% dropout is applied to the 11-way softmax output layer.

$\cdot$ \textit{Multi-layer} architecture's first layer has the same settings as \textit{single-layer} but it is deepen by two convolutional layers of 128 filters of size $3{\times}3$, one fully-connected layer of size $256$ and a 11-way softmax output layer. 50\% dropout is applied to all the dense layers and 25\% for convolutional layers. Each convolutional layer is followed by max-pooling: first wide layer - \textit{MP(12,16)}; deep layers - \textit{MP(2,2)}. 

Implementations are available online\footnote{https://github.com/Veleslavia/EUSIPCO2017}. STFT is computed using 512 points FFT with a hop size of 256. Audios where down-sampled to 12kHz. Each convolutional layer is followed by batch normalization \cite{ioffe2015batch}. All convolutions use \textit{same} padding. Therefore, the dimensions of the feature maps out of the first convolutional layer are still equivalent to the input -- time and frequency. Then, the resulting feature map of the \textit{MP(12,16)} layer can be interpreted as an eight-bands summary (96/12=8). This max-pool layer was designed considering: \textit{(i)} is relevant to know in which band a given filter shape is mostly activated -- as a proxy for knowing in which pitch range timbre is occurring; and \textit{(ii)} is not so relevant to know when it is mostly activated. To obtain instrument predictions from the softmax layer we use the same strategy as Han \textit{et al.} \cite{han2017deep}: estimations for the same song are averaged and then a threshold of $0.2$ is applied. In Table~\ref{irmas_results} we report the standard metrics for this dataset such as micro- and macro- precision, recall and f-beta score (f1). The micro- metrics are calculated globally for all testing examples while the macro-metrics are calculated label-wise and the unweighted average is reported. 

\begin{table}[h]
\caption{Recognition performance for \textit{IRMAS} dataset.}
\label{irmas_results}
\begin{tabular}{l|ccc|ccc}
\toprule
{} & \multicolumn{3}{|c|}{Micro} & \multicolumn{3}{|c}{Macro}\\
Model / \#param & P & R & F1 & P & R & F1 \\
\midrule
Bosch \textit{et al.} & 0.504 & 0.501 & 0.503 & 0.41 & 0.455 & 0.432 \\
Han \textit{et al.} / 1446k & \textbf{0.655} & \textbf{0.557} & \textbf{0.602} & 0.541 & 0.508 & 0.503 \\
Single-layer / 62k & 0.611 & 0.516 & 0.559 & 0.523 & 0.480 & 0.484 \\
Multi-layer / 743k & 0.650 & 0.538 & 0.589 & \textbf{0.550} & \textbf{0.525} & \textbf{0.516} \\
\bottomrule
\end{tabular}
\end{table}

\textit{Multi-layer} achieved similar results as the state-of-the-art with twice fewer \textit{\#param}. This result denotes how efficient are the proposed architectures. Moreover, note that small filters are also used within the proposed architecture. We found these filters to be important for achieving state-of-the-art performance -- although no instruments with such \textit{small} time-frequency signature (such as kick drum sounds or bass notes) are present in the dataset. However if \textit{m=5} filters are substituted with \textit{m=50} filters, the performance does not drop dramatically. Finally note that \textit{single-layer} still achieves remarkable results: it outperforms the standard bag-of-frames + SVM approach.

\subsection{Music auto-tagging}
Automatic tagging is a multi-label classification task. We approach this problem by means of the MagnaTagATune dataset \cite{law2009evaluation} -- with 25.856 clips of $\approx$ 30 seconds sampled at 16kHz. Predicting the top-50 tags of this dataset (instruments, genres and others) has been a popular benchmark for comparing deep learning architectures \cite{choi2016automatic}\cite{dieleman2014end}. Architectures from Choi \textit{et al.} \cite{choi2016automatic} and Dieleman \textit{et al.} \cite{dieleman2014end} are set as baselines -- that are state-of-the-art examples of architectures based on small-rectangular filters and high filters, respectively. Therefore, this dataset provides a nice opportunity to explore the trade off between leaning little context with small-rectangular filters and risking to fit noise with high filters. Choi \textit{et al.}'s architecture consists of a CNN of five layers where filters are of size $3{\times}3$ with an input of size $96{\times}187$. After every CNN layer, batch normalization and max-pool is applied. Dieleman \textit{et al.}'s architecture has two CNN layers with filters of $M{\times}8$ and $M'{\times}8$ size, respectively. The input is of size $128{\times}187$. After every CNN layer a max-pool layer of $1{\times}4$ is applied. Later, the penultimate layer is a fully connected layer of 100 units. An additional baseline is provided: \textit{Small-rectangular}, which is an adaption of Choi \textit{et al.}'s architecture to have the same input and number of parameters as Dieleman \textit{et al.} All models use a 50-way sigmoidal output layer and STFT was performed using 512 points FFT with a hop size of 256. 
\begin{table}[h]
\centering
\caption{Models performance for MagnaTagATune dataset.}
\label{resultsAT}
\begin{tabular}{ll|ll}
\toprule
Model & AUC/\#param & Model & AUC/\#param \\
\midrule
Small-rectangular  & 0.865 / 75k & Choi \textit{et al.} \cite{choi2016automatic} &  \textbf{0.894} / 22M\footnotemark\\
Dieleman \textit{et al.} \cite{dieleman2014end} & 0.881 / 75k  & Proposed x2 & \textbf{0.893} / 191k \\
Proposed & 0.889 / 75k& Proposed x4 & 0.887 / 565k \\
\bottomrule
\end{tabular}
\end{table}
\footnotetext{{Although equivalent results can be achieved with 750k parameters.}}

Our experiments reproduce the same conditions as in Dieleman \textit{et al.} since the proposed model adapts their architecture to the proposed design strategy --
we uniquely modify the first layer to have many musically motivated filter shapes. Other layers are kept intact. This allows to isolate our experiments from confounding factors, so that we uniquely measure the impact of increasing the representational capacity of the first layer. Inputs are set to be of size $128{\times}187$ -- since input spectrograms ($\approx$3 seconds) are shorter than the total length of the song, estimations for the same song are averaged.
We consider the following frequency contexts as relevant for this dataset: \textit{m=100} and \textit{m=75} to capture different wide spectral shapes (\textit{e.g.} genres timbre or guitar), and \textit{m=25} to capture shallow spectral shapes (\textit{e.g.} drums). For consistency with Dieleman \textit{et al.}, we consider the following temporal context: \textit{n=}[\textit{1,3,5,7}]. We use several filters per shape in the first layer:

$\cdot$ \underline{\textit{m=100}}: 10x $100{\times}1$, 6x $100{\times}3$, 3x  $100{\times}5$ and 3x $100{\times}7$.

$\cdot$ \underline{\textit{m=75}}: 15x $75{\times}1$, 10x $75{\times}3$, 5x  $75{\times}5$ and 5x $75{\times}7$. 

$\cdot$ \underline{\textit{m=25}}: 15x $25{\times}1$, 10x $25{\times}3$, 5x  $25{\times}5$ and 5x $25{\times}7$. 

For merging the resulting feature maps, these need to be of the same dimension. We zero-pad the temporal dimension before first layer convolutions and use max-pool layers: \textit{MP(M',4)} -- note that all resulting feature maps have the same dimension: $1{\times}N'$, and are pitch invariant. 
50\% dropout is applied to all dense layers. We also evaluate variants of the proposed model where the number of filters per shape in the first layer are increased according to a factor -- other layers are kept intact. Implementations are available online\footnote{https://github.com/jordipons/EUSIPCO2017}.

We use area under the ROC curve (AUC) as metric for our experiments. Table \ref{resultsAT} (left column) shows the results of three different architectures with the same number of parameters. The proposed model outperforms others, denoting that architectures based on the design strategy we propose can better use the capacity of the network. Moreover, Table \ref{resultsAT} (right column) shows that is beneficial to increase the representational capacity of the first layer -- up to the point where we achieve equivalent results to the state-of-the-art while significantly reducing the \textit{\#param} of the model.

\section{Conclusions}

Inspired by the fact that it is hard to identify the adequate combination of parameters for a deep learning model --which leads to architectures being difficult to interpret--, we decided to incorporate domain knowledge during the architectural design process. This lead us to discuss some common practices when designing CNNs for music classification -- with a specific focus on how to learn timbre representations. This discussion motivated the design strategy we present for modeling timbre using spectroram-based CNNs. Several ideas were proposed to achieve pitch, loudness, duration and spatial position invariance with CNNs. Moreover, we proposed actions to increase the efficiency of these models. The idea is to use different filter shapes in the first layer that are motivated by domain knowledge -- namely, using different musically  motivated filter shapes in the first layer.
A part from providing theoretical discussion and background for the proposed design strategy, we also validated it empirically. Several experiments in three datasets for different tasks related to timbre (singing voice phoneme classification, musical instrument recognition and music auto-tagging) provide empirical evidence that this approach is powerful and promising. 
In these experiments, we evaluate several architectures based on the presented design strategy -- that has proven to be very effective in all cases. These results support the idea that increasing the representational capacity of the layers can by achieved by using different filter shapes.
Specifically, proposed architectures used several filter shapes having the capacity of capturing timbre with \textit{high enough} filters. 
Moreover, we found very remarkable the results of the proposed single-layer architectures. Since single-layer architectures use a reduced amount of parameters, these might be very useful in scenarios where small data and a few hardware resources are available. Furthermore, when deepen the network we were able to achieve equivalent results to the state-of-the-art -- if not better. 
As future work we plan to relate these findings with previous research (where a similar analysis was done for designing CNNs for modeling temporal features \cite{ponsdesigning1}), to extend this work to non-musical tasks, and to inspect what filters are learning.

\section*{Acknowledgments}

We are grateful for the GPUs donated by NVIDIA. This work is partially supported by: the Maria de Maeztu Units of Excellence Programme (MDM-2015-0502), the CompMusic project (ERC grant agreement 267583) and the CASAS Spanish research project (TIN2015-70816-R). Also infinite thanks to E. Fonseca and S. Oramas for their help.

\bibliographystyle{IEEEtran}
\bibliography{references}{}

\end{document}